\documentclass[a4paper,12pt]{article}
\usepackage{etex}
\usepackage{epsf}
\usepackage[dvips]{graphicx}
\usepackage{a4wide}
\usepackage{amsmath}
\usepackage{amscd}
\usepackage{amsfonts}
\usepackage{natbib}
\usepackage{multirow}

\usepackage{microtype} % Slightly tweak font spacing for aesthetics
\usepackage[hang, small,labelfont=bf,up,textfont=it,up,skip=2.5pt]{caption}
\usepackage{booktabs}
\usepackage{lettrine}

\usepackage{abstract} % Allows abstract customization
\usepackage{titlesec} % Allows customization of titles
\titleformat{\section}[block]{\large\scshape\centering}{\thesection.}{1em}{} % Change the look of the section titles
\titleformat{\subsection}[block]{\large}{\thesubsection.}{1em}{} % Change the look of the section titles

\usepackage{titling} 
\usepackage{amssymb}
\usepackage{caption,subfig}
\usepackage{longtable}
\usepackage{rotating}
\usepackage{wrapfig}
\usepackage{pdflscape}
\usepackage{color}
\usepackage{multirow}
\usepackage{rotating}
\usepackage{natbib}
\usepackage{enumerate}
\usepackage{enumitem}
\usepackage{dsfont}
\usepackage{setspace}
\usepackage{anysize}
\usepackage{booktabs}
\usepackage{soul}
\usepackage{float}
\usepackage{enumitem}
\usepackage{tikz}
\usepackage{pifont}
\usepackage[hidelinks]{hyperref}
\usepackage{moresize}
\usetikzlibrary{arrows}
\usepackage{multirow}
\usepackage{ntheorem}
\usepackage{geometry}
\usepackage{tabularx}
\hypersetup{colorlinks=true,urlcolor=blue,linkcolor=blue,citecolor=blue}
\usepackage[section]{placeins}
\usepackage{textcomp}

\newtheorem{proposition}{Proposition}
\newtheorem{corollary}{Corollary}
\newtheorem{algorithm}{Algorithm}

\newtheorem{remark}{Remark}

\newcommand{\qed}{\nobreak \ifvmode \relax \else
      \ifdim\lastskip<1.5em \hskip-\lastskip
      \hskip1.5em plus0em minus1/2em \fi \nobreak
      \vrule height0.75em width1/2em depth0.25em\fi}
      
\def\E{{\mathbb E}}

\def\R{{\mathbb R}}

\def\dim{{\rm dim}}

\def\bb1{1}

\def\bb{{\bf b}}
      
\marginsize{3.18cm}{3.18cm}{3.18cm}{3.18cm}
\long\def\symbolfootnote[#1]#2{\begingroup
\def\thefootnote{\fnsymbol{footnote}}\footnote[#1]{#2}\endgroup}

\DeclareGraphicsExtensions{.eps}
\date{}

\title{Fast, Robust Inference for Linear Instrumental Variables Models using Self-Normalized Moments}
\author{Eric Gautier (Toulouse School of Economics), \\
Christiern Rose (University of Queensland)\thanks{Following editorial advice, this paper presents and extends a subset of the analyses originally conducted in \cite{gautier18} (see Section 3.3). We thank James Stock, Elie Tamer, Ulrich M{\"u}ller, and three anonymous referees for their suggestions. We warmly thank Alexandre Tsybakov whose reflections on this project have been fundamental. We acknowledge financial support from ERC grant POEMH, ANR-17-EURE-0010 and ARC grant DP220101043. }}

\renewcommand{\maketitlehookd}{%
\begin{abstract}
\noindent We propose and implement an approach to inference in linear instrumental variables models which is simultaneously robust and computationally tractable. Inference is based on self-normalization of sample moment conditions, and allows for (but does not require) many (relative to the sample size), weak, potentially invalid (i.e., the exclusion restriction fails) or potentially endogenous instruments, as well as for many regressors and conditional heteroskedasticity. Our coverage results are uniform and can deliver a small sample guarantee. We develop a new computational approach based on semidefinite programming, which we show can equally be applied to rapidly invert existing tests (AR, LM, CLR, etc.).\\

\noindent\textbf{Key words:} High-dimensional inference, weak instruments, many instruments, invalid instruments, endogenous instruments.
\end{abstract}
}

\begin{document}

%NOTES
%1) Could put BSOS or s-BSOS in the appendix and use in sims (decide later). Or cite the version of our revision which uses them.
%2) Application to `bad controls'?
%3) Partially identified models. Another paper? Ask Christian B if there is a contribution and to co-author?

\maketitle

\section{Introduction}

Instrumental variables are widely used in applied econometrics, yet computationally tractable robust inference remains challenging. It is well known that inference robust to weak instruments can be conducted by inverting robust tests. %Examples include the Anderson Rubin (AR) \citep{anderson49}, Lagrange-multiplier (LM) \citep{kleibergen02,moreira02} and Conditional Likelihood Ratio (CLR) \citep{moreira03} tests. 
However, to our knowledge there do not exist inference methods which are simultaneously robust to weak instruments, many instruments/regressors (e.g., larger than the sample size), potentially invalid instruments (i.e., an exclusion restriction may be violated) and potentially endogenous instruments, nor which can offer coverage guarantees in small samples. Moreover, inverting robust tests can be computationally challenging \citep{andrews19} because their non-rejection regions are not convex \citep{mikusheva10} and a grid search is infeasible with just a handful of regressors (see \cite{andrews16con}, supplementary material).

We address both the statistical and computational challenges. First, we provide an approach to inference based on self-normalized sample moment conditions, which we refer to as Self-Normalized Instrumental Variables (SNIV). Due to the minimal assumptions it requires, SNIV simultaneously allows for weak instruments and conditional heteroskedasticity, and can be applied equally to the standard low-dimensional setting (large sample size, few regressors, few instruments) and to the high-dimensional setting in which the number of instruments and/or regressors can be large, possibly much larger than the sample size. For example, in a model with a single endogenous regressor, SNIV could be used to construct a confidence interval which is simultaneously robust to many and to weak instruments.\footnote{To further motivate our framework, there are several reasons to expect models with multiple endogenous regressors to become increasingly popular. Potential applications include demand systems with many goods and endogenous expenditure \citep{gautier21} or prices \citep{belloni22}, as well as models of peer effects with unknown peer relationships \citep{rose18}. More broadly, due to increasing availability of rich datasets and the potential to allow for heterogeneous treatment effects by using interactions of the treatment with individual characteristics, applied research has recently considered models with multiple exogenous variables of interest (e.g., \cite{farrell20}). A natural extensionis to allow for endogenous treatment \citep{belloni22}.} We extend SNIV to settings in which one or more instrument may be `invalid' (i.e., the exclusion restriction fails, see \cite{kolesar15,kang16}) or endogenous without requiring a pre-test, and propose the use of an a-priori upper bound on the number of invalid/endogenous instruments for settings in which the set of identifiable parameters would otherwise be unbounded.

Second, we provide a computational implementation of SNIV which we show can also be applied to rapidly invert other robust tests. This is because test inversion can typically be cast as a semi-algebraic optimization problem, hence we can apply methods from the literature on semi-algebraic optimization. The basic idea to deal with computational intractability is to attempt to solve a hiearchy of semidefinite optimization problems. Solving each optimization problem delivers an outer bound on the confidence region. As we proceed up the hierarchy, the bounds become sharper at the expense of greater computational burden. This allows the researcher to effectively trade off sharpness with available computational resources. In practice, the bounds obtained towards the beginning of the hierarchy are often exact. A simple diagnostic informs the researcher if exact bounds have been attained. Similar computational approaches have been applied by \cite{gautier18}, \cite{lee20} and \cite{auerbach22}.

In contrast to a grid search, our approach can be applied to settings with multiple regressors. In contrast to heuristic/local optimization methods, our approach guarantees an outer bound on the confidence region, hence does not risk compromising the coverage guarantee. We illustrate how our computational approach can be used to rapidly invert existing robust tests by combining it with the results of \cite{guggenberger12}, \cite{guggenberger19} and \cite{guggenberger21} to obtain weak instrument robust Anderson-Rubin (AR) confidence intervals which can be computed near instantaneously, even when a grid search is infeasible. We also show that our approach can be applied to rapidly invert other robust tests such as the Lagrange-multiplier (LM) test \citep{kleibergen02,moreira02} and the Conditional Likelihood Ratio (CLR) test \citep{moreira03}. 

We conduct a Monte-Carlo experiment in which we demonstrate SNIV and AR confidence regions are both easily implemented in a setting in which a grid search is computationally intractable. SNIV has similar coverage to the AR test in designs with either strong or weak instruments. With many instruments, SNIV maintains coverage close to the nominal level but AR does not. We also show that SNIV can be applied to conduct informative inference with invalid instruments and endogenous instruments in challenging designs, to which existing approaches cannot be applied.

\subsection{Related literature}

Our work is related to the literature on many instruments (e.g., \cite{bekker94,angrist99,donald01,anderson05,chao05,stock05,hansen08,ackerberg09,van10,chao12,hausman12,anatolyev13,hansen14,kolesar18}; see \cite{anatolyev19} for a recent review) and weak instruments (e.g., \cite{anderson49,kleibergen02,moreira02,moreira03,mikusheva10,guggenberger12,andrews16con,guggenberger19,guggenberger21}; see \cite{andrews19} for a recent review), but allows simultaneously for weak instruments and for the number of regressors and/or instruments to be large, possibly much larger than the sample size. This is because we conduct inference based on moderate deviations of self-normalized sample moments (e.g., \cite{pinelis94,bertail08,jing03}) instead of using a Central Limit Theorem.

The most closely related papers are \cite{gautier11}, \cite{belloni12}, \cite{gold20}, \cite{gautier21} and \cite{belloni22}, all of which also consider the linear instrumental variables model in a potentially high-dimensional setting. \cite{gautier11} and \cite{gautier21} suggest to combine a point estimator with lower bounds on its sensitivity characteristics to perform robust inference, but their confidence region is larger than ours and the authors do not provide a disciplined way to trade off computational complexity and sharpness when implementing their approach. \cite{belloni12} discuss inference based on self-normalization, but do not propose a practical computational solution, nor allow for invalid/endogenous instruments. \cite{gold20}, \cite{gautier21} and \cite{belloni22} propose confidence regions for a subset of parameters of interest (e.g., confidence intervals) but these rely on stronger assumptions than we use below. For example, these papers propose methods which are not robust to weak instruments and do not allow for potentially invalid nor endogenous instruments. We view our work as complementary to \cite{gold20}, \cite{gautier21} and \cite{belloni22}, providing the applied researcher with a more robust alternative, but one which may sometimes yield wider bounds in practice.

Finally, our extensions of SNIV are related to the literature on invalid and endogenous instruments. Regarding invalid instruments (e.g., \cite{kolesar15,kang16}), we allow for a setting with multiple endogenous regressors, potentially weak instruments, and for the number of instruments and/or regressors to be larger than the sample size. Regarding endogenous instruments  (e.g., \cite{sargan58,hansen82,anatolyev11,lee12,chao14}), we do not perform specification tests, but instead perform inference directly, accounting for potential endogeneity of an unknown subset of instruments.

We proceed as follows. Section \ref{model} sets out our model. Section \ref{sniv} defines SNIV, establishes its coverage guarantee, and provides extensions to potentially invalid and endogenous instruments. Section \ref{comp} presents our computational method, which we apply to invert existing robust tests in Section \ref{classic}. Section \ref{monte} presents a Monte-Carlo experiment and Section \ref{conc} concludes. All proofs are gathered in the appendix.

%\begin{itemize}
%	\item SNIV sets work for many, few, weak and endogenous IVs and many regressors. Uniform and finite sample coverage.
%	\item Computationally feasible method. 
%	\item Computational method can also be applied to other settings, e.g., classical IV models (section 5) and partial identification (section 6, if kept). Hence we can now allow researchers to rapidly compute e.g., AR confidence intervals, which was previously impossible when $d_X$ was larger than a handful
%	\item Monte-Carlo: SNIV works as well as AR in classical settings, but is much more broadly applicable. Fast computation in a setting where gridding is impossible.
%\end{itemize}

\subsection{Setup \& notation}
To simplify the exposition we consider an i.i.d. sample of size $n$. The i.i.d. setting is not critical for our results and can be relaxed by using an alternative choice of $r_n$ below, for which we provide appropriate references. The population model comprises an outcome $Y$, regressors $X\in\R^{d_X}$, and instrumental variables $Z\in\R^{d_Z}$ of joint distribution $\mathbb{P}$. $\E$ is the expectation under $\mathbb{P}$ and $\E_n$ is its sample counterpart. Our results apply to a sequence of models indexed by $n$. For simplicity of exposition we do not make this explicit, but we occasionally note that certain objects can depend on $n$. To allow for high-dimensional data, the relative magnitudes of $n,d_X$ and $d_Z$ are unrestricted, and both $d_X$ and $d_Z$ can grow with $n$. %For a mean zero random variable $A$, $\sigma_{A}\triangleq\E[A^2]^{1/2}$. We denote stacked matrices in bold, e.g.,  
%$\bold{Y}\in\R^n$ and 
%$\bold{X}\in\mathcal{M}_{n,d_X}$, where $\mathcal{M}_{{d},{d'}}$ is the set of ${d}\times {d'}$ matrices. 
%For %$d\in\N$ and 
%$d\in\mathbb{N}$ and a random vector $W\in\R^{d}$, $\E_n[W]$ is the sample mean. % and, 
%the sample and population means are  $\E_n[R]\triangleq\sum_{i\in[n]} R_{i,\cdot}/n$ and $\E[R]\triangleq\sum_{i\in[n]} \E\left[R_{i,\cdot}\right]/n$ and
%for $k\in[d]$ and $p\ge0$, $\E_n\left[W_k^p\right]$ is obtained by replacing $W_{k}$ by $W_{k}^p$. 
%$\bold{D}_{\bold{W}}$ is the diagonal matrix with entries 
%$\mathbb{E}_n\left[W_k^2\right]^{-1/2}$ for $k\in[d]\triangleq \{1,2,...,d\}$, and $D_W$ its population counterpart. 
For %the function 
$b\in\R^{d_X}$, %\to 
$U(b)\triangleq Y-X^{\top}b$,  % and its $k$th component is denoted by $r_{k,i}$. %Conditional expectations with non i.d. data is defined like population means. 
%When $r_n$ is a matrix, the sample and population means are defined entrywise.  %When $(a_i)_{i\in[n]}$ are random vectors, $\bold{A}$ is the matrix where each row is $a_i^{\top}$ for $i\in[n]$.
%For $b\in\R^{d_X}$, 
 $\mathbb{P}(b)$ is the distribution of $\left(X,Z,U(b)\right)$ implied by $\mathbb{P}$. %, and $\widehat{\sigma}(b)^2\triangleq\E_n[U(b)^2]$.
%We use the diagonal matrices $\bold{D}_{\bold{X}}$, $\bold{D}_{\bold{Z}}$ of entries $\left(\bold{D}_{\bold{X}}\right)_{k,k}\triangleq\mathbb{E}_n\left[X_k^2\right]^{-1/2}$ for $k\in[{d_X}]$ and $\left(\bold{D}_{\bold{Z}}\right)_{l,l}\triangleq\mathbb{E}_n\left[Z_l^2\right]^{-1/2}$ for $l\in[{d_Z}]$. % to scale $\bold{X}$ and $\bold{Z}$. %They are diagonal ${d_X}\times {d_X}$ and ${d_Z}\times {d_Z}$ matrices. 
%We write $D_Z$ and $ D_X$ for the population counterparts, 
%We write $\widehat{\Psi}\triangleq \bold{D}_{\bold{Z}}\E_n[ZX^{\top}]\bold{D}_{\bold{X}}$ and $\Psi\triangleq D_Z\E[ZX^{\top}]D_X$. 
%The set $S_I\subseteq [{d_X}]$ collects the indices of the regressors which are also instruments 
%(if the researcher is uncertain about a regressor being exogenous he/she should avoid including the corresponding index in $S_I$) 
%and $S_Q\subseteq [{d_X}]$ of size $d_Q$ collects those of the regressors of questionable relevance. 
%When we make inference on a vector of functions,  %$\varPhi \beta$, 
%its dimension is $d_{\varPhi}$. 
%Some results are asymptotic in $n\to\infty$ in which case %, though we do not make the dependence explicit, 
%${d_Z}$, ${d_X}$, $d_{\varPhi}$ and $d_Q$ can increase with $n$ and triangular arrays are permitted.  
%Inequality between vectors is entrywise.   %$M_{k,\cdot}$ (resp. $M_{\cdot,k}$) is the $k^{\rm th}$ row (resp. column) of $M$. %$S\subseteq [d]$ and 
%$\indic$ is the indicator function. 
For $S\subseteq [d]\triangleq \{1,2,...,d\}$, $|S|$ is its cardinality and $S^c$ its complement. For $\Delta\in\mathbb{R}^d$, $S(\Delta)\triangleq\{k\in[d]:\ \Delta_{k}\ne0\}$ and %and  %be the support of $\Delta$,  %
%$\Delta_S\triangleq\left(\Delta_{k}\indic{\{k\in S\}}\right)_{k\in[d]}$. % for $S\subseteq [d]$. 
$|\Delta|_p$ is the $\ell^p$-norm of %a vector 
$\Delta$. % or a vectorization if $\Delta$ is a matrix.
For a polynomial $p$, deg$(p)$ is its degree. We use $\mathbf{M}\succcurlyeq \mathbf{0}$ to say that the matrix $\mathbf{M}$ is positive semidefinite.

\section{Model}\label{model}

The linear instrumental variables model is%\vspace{-.2cm}
\begin{align}
&\E\left[ZU(\beta)\right]=0,\label{einstr}\\
&\beta\in\mathcal{B},\ \mathbb{P}(\beta)\in\mathcal{P},\label{econstraints}
\end{align}
where $\mathcal{B}\subseteq \mathbb{R}^{d_X}$ is the parameter space and $\mathcal{P}$ is a nonparametric class. %e.g.,\\ 
%{\bf Class 1:} 
%\emph{$Z_lU(\beta)$ is %)_{i\in[n]}$ is %i.i.d. and 
%symmetric for all $l\in[d_Z]$ and ${d_Z}<9\alpha/\left(4e^3\Phi\left(-\sqrt{n}\right)\right)$}\\ 
%where $\alpha\in(0,1)$ is a confidence level and $\Phi$ the normal CDF. Other classes allow for non i.d. and dependent data, and asymmetry (see Section \ref{cs3}). Their basic versions do not restrict the joint distribution of $Z$ and $X$. 
%All but Class 4 allow for conditional heteroscedasticity.  
The set $\mathcal{I}$ collects the vectors which satisfy \eqref{einstr}-\eqref{econstraints}. As will be made clear below, our results are for all $\beta\in\mathcal{I}$, hence for the true value $\beta^*$. % when we do not consider a triangular array of models.
%They place no upper bound on ${d_X}$ and $d_Q$.
We use the class $\mathcal{P}$ to permit the use of results on moderate deviations of self-normalized sums for inference. We consider four classes, including\\

\noindent {\bf Class 1.} 
\emph{%$(\bold{Z}_{i,\cdot}^{\top}\bold{U}_{i}(\beta))_{i\in[n]}$ are independent, 
There exists $\delta$ in $(0,1]$ and $\mu_{2+\delta}>0$ such that
$$\displaystyle\left|\left(
\left(\E\left[|Z_lU(\beta)|^{2+\delta}\right]\right)
\left(\E\left[Z_l^2U(\beta)^2\right]\right)^{-(2+\delta)/2}\right)_{l\in[{d_Z}]}\right|_{\infty}
\le\mu_{2+\delta},$$ 
and
${d_Z}\le \alpha/(2\Phi(-n^{1/2-1/(2+\delta)}\mu_{2+\delta}^{-1/(2+\delta)}))$},

\noindent where $\alpha\in(0,1)$ is a confidence level and $\Phi$ the normal CDF;\\

\noindent {\bf Class 2.} 
\emph{%$(\bold{Z}_{i,\cdot}\bold{U}_{i}(\beta))_{i\in[n]}$ are i.i.d., 
$\exists \mu_4>0:$ %such that
$\max_{l\in[{d_Z}]}\mathbb{E}[Z_l^4U(\beta)^4](\mathbb{E}[Z_l^2U(\beta)^2])^{-2}\le \mu_{4}$, ${d_Z}<\alpha\exp\left(n/\mu_4\right)/(2e+1)$ and $n-\mu_4\log({d_Z}(2e+1)/\alpha)\geq n/2$};\\ 

%For Class 2 when $n-\delta_4\log({d_Z}(2e+1)/\alpha)\ge n/2$ one can take $r_n=2\sqrt{\log({d_Z}(2e+1)/\alpha)/n}$. 
%\cite{BGHK1} propose an approach to estimate $\delta_4$.\\ 
%{\bf Class 3:} 
%\emph{$(\bold{U}_{i}(\beta))_{i\in[n]}$ are independent and symmetric conditional on $\bold{Z}$ or 
%$(\bold{Z}_{i,\cdot}^{\top}\bold{U}_{i}(\beta))_{i\in[n]}$ are independent and symmetric; \begin{equation*}
%r_n=\sqrt{\frac{2\log(2{d_Z}/\alpha)}{n}}.
%\end{equation*}}

\noindent{\bf Class 3.} 
\emph{$Z_lU(\beta)$ is %)_{i\in[n]}$ is %i.i.d. and 
symmetric for all $l\in[d_Z]$ and ${d_Z}<9\alpha/\left(4e^3\Phi\left(-\sqrt{n}\right)\right)$}.\\

\noindent Classes 1-2 require mild bounds on ratios of moments, whereas Class 3 requires no bounds but uses symmetry. Further classes allowing for dependence and non i.d. data can be found in \cite{chen16} and references therein.  In Section \ref{mboot} we consider a fourth class based on Gaussian approximation rather than self-normalization. %We require no further assumptions. 

\begin{remark}\label{r1}
The model in \eqref{einstr}-\eqref{econstraints} can be obtained by first partialling-out a low-dimensional vector of exogenous regressors. To simplify the exposition we do not make this explicit.
\end{remark}

\section{Self Normalized Instrumental Variables}\label{sniv}
The $1-\alpha$ SNIV confidence set is
\begin{align}
	\widehat{\mathcal{C}}\triangleq\left\{\beta\in\mathcal{B}: |\mathbf{D}(\beta)\E_n[ZU(\beta)]|_\infty\leq r_n \right\},\label{sniv}
\end{align}
where $\mathbf{D}(\beta)$ is the $d_Z\times d_Z$ positive, diagonal matrix with $l^{\text{th}}$ diagonal element $\mathbb{E}_n[Z_l^2U(\beta)^2]^{-1/2}$ used to self-normalize the $d_Z$ moments and $r_n$ depends on the class. Under Class 1 we set $r_n=-\Phi^{-1}\left(\alpha/(2{d_Z})\right)/\sqrt{n}$. Under Class 2 we set   $r_n=2\sqrt{\log({d_Z}(2e+1)/\alpha)/n}$. Under Class 3 we set $r_n=-\Phi^{-1}(9\alpha/(4d_Ze^3))/\sqrt{n}.$

\begin{proposition}\label{propcover}
Consider the model in \eqref{einstr}-\eqref{econstraints}. If $\mathcal{P}$ is Class 1 we have
\begin{align}
\lim_{n\to\infty}\inf_{(\beta,\mathbb{P}):\beta\in\mathcal{I}}\mathbb{P}(\beta\in\widehat{\mathcal{C}})\geq 1-\alpha\label{cover}.
\end{align}
If $\mathcal{P}$ is either Class 2 or Class 3 we have
\begin{align}
\inf_{(\beta,\mathbb{P}):\beta\in\mathcal{I}}\mathbb{P}(\beta\in\widehat{\mathcal{C}})\geq 1-\alpha\label{cover}.
\end{align}

\end{proposition}
Proposition \ref{propcover} shows that the coverage of SNIV is at least the nominal level uniformly over the identifiable parameters and the distributions of the data they imply. Beyond the class used, no further assumptions are needed. Classes 1-3 allow for conditional heteroscedasticity, do not restrict the joint distribution of $X$ and $Z$ (hence are robust to weak instruments) and have very mild requirements on the relative magnitudes of $n$ and $d_Z$ (hence are robust to many regressors and/or instruments). For Class 1 the coverage guarantee is asymptotic in $n$ in such a way that $d_X$ and $d_Z$ can grow with (and be much larger than) $n$.  For Classes 2-3 the coverage guarantee is for any $n$. 

The SNIV confidence set collects vectors for which the $\ell_\infty$ deviation from zero of the self-normalized sample moment is at most $r_n$. The core components which deliver uniformity, finite sample validity and robustness to identification are the $\ell_\infty$-norm and self-normalization of the moments. The $\ell_\infty$-norm is crucial so as to allow for $d_Z$ larger than $n$ because it permits $r_n$ to be of the order $\log(d_Z)/\sqrt{n}$. This means that the SNIV confidence set can be small even when the number of instruments is much larger than the sample size.\footnote{For Class 3, $r_n\le2\log\left(4{d_Z}e^3/(9\alpha)\right)/\sqrt{n},$ $\forall\alpha\in[0,1],d_Z\ge1$(because $\Phi^{-1}(a)\ge2\log(a)$ if $0<a\le\exp(-1/(4\pi))$).} Note also that $d_X$ can be arbitrarily large with respect to $n$. 

%In sum, SNIV is robust to identification, can be used in small or large samples with or without conditional heteroscedasticity, and can be applied when the number of instruments and/or regressors is much larger than the sample size. %Self-normalization allows us to apply the results of \cite{pinelis94,bertail08} or \cite{jing03} (among others) to choose $r_n$ which delivers \eqref{cover}.

\begin{remark} If $\mathcal{B}$ is defined by polynomial (in)equalities of degree at most 2, the SNIV confidence set is defined by polynomial inequalities of degree at most 2 (we show this in Proposition \ref{propsnivg}), so it can be empty, unbounded or disconnected depending on the (random) values of the polynomial coefficients. Possible unboundedness is unavoidable for confidence sets which are robust to identification \citep{dufour97}.
\end{remark}

\subsection{Gaussian approximation}\label{mboot}

%\textbf{[What is written below comes from the STIV paper (v4). We may prefer to use the class from the most recent revision (commented out below), but it leads to a different $\widehat{r}$ which depends on an unknown sequence ($\zeta_n$), which depends on $\tau_n$ etc.. So it isn't clear how it could be used in practice. We could also just drop this section and use Classes 1-3]}

The SNIV confidence set may be conservative when the instruments are strongly correlated with one another because $r_n$ is based on a union bound over $d_Z$ self-normalized sample moments. \cite{gautier21} propose an alternative to self-normalization based around the multiplier bootstrap of \cite{chernozhukov13}, which we implement here. We modify the SNIV confidence set by replacing $\mathbf{D}(\beta)$ by $\E_n[U(\beta)]^{-1/2}\mathbf{D_Z}$ and $r_n$ by the $1-\alpha$ quantile of $|\mathbf{D_Z}\E_n[ZW]|_\infty$ (computed by simulation) in its definition, where $\mathbf{D_Z}$ is a $d_Z\times d_Z$ diagonal matrix with $l^{th}$ diagonal element $\E_n[Z_l^2]^{-1/2}$ and $W$ is a standard normal random variable which is independent of $Z$. The corresponding class is\\

\noindent {\bf Class 4.}  There exist constants $C$ and $c$, and $B_n$ such that, for all $(\beta,\mathbb{P})$: $\beta\in\mathcal{I}$, $U(\beta)\perp Z$; $|Z|_\infty\leq B_n$ (a.s.); $\E[U(\beta)^4]\leq C$; and $B_n^4\log(d_Zn)^7/n\leq Cn^{-c}$,\\

\noindent which delivers the following coverage guarantee. 

\begin{proposition}\label{propcovermboot}
Consider the model in \eqref{einstr}-\eqref{econstraints}. If $\mathcal{P}$ is Class 4, then, for $\widehat{\mathcal{C}}$ based on the multiplier bootstrap, we have
\begin{align}
\lim_{n\to\infty}\inf_{(\beta,\mathbb{P}):\beta\in\mathcal{I}}\mathbb{P}(\beta\in\widehat{\mathcal{C}})\geq 1-\alpha\label{coveras}.
\end{align}
\end{proposition}
Further classes based on Gaussian approximation but allowing for non i.d. and dependent data can be found in \cite{zhang17} and references therein.

%\noindent {\bf Class 4:} ${d_Z}\ge3$, $M_{U}, M_Z,q_2>0$, and a sequence $(B_n)_{n\in\N}$ such that $B_n\ge1$. For all $(\beta,\mathbb{P})$: $\beta\in\mathcal{I}$ and $q_1\in[2]$, %we have
%\begin{enumerate}
%\item$\mathbb{E}\left[\left.U(\beta)^2\right|Z\right]=\mathbb{E}\left[U(\beta)^2\right]$;
%\item $\mathbb{E}\left[ \left(U(\beta)^2/\mathbb{E}[U(\beta)^2]-1\right)^2 \right]\leq M_U$, $\E\left[\left|D_Z\left(ZZ^{\top}-\E\left[ZZ^{\top}\right]\right)D_Z\right|_{\infty}^2\right]\le M_Z$;
%\item$\left|\left(\max\left(\E\left[\left(\E[Z_l^2]^{-1/2}Z_lU(\beta)\E[U(\beta)^2]^{-1/2}\right)^{2+q_1}\right],
%\E\left[\left(\E[Z_l^2]^{-1/2}Z_lW\right)^{2+q_1}\right]
%\right)\right)_{l=1}^{d_Z}\right|_{\infty}
%\le B_n^{q_1}$;
%\item$\max\left(\E\left[\left(\left|D_ZZU(\beta)\right|_{\infty}/(B_n\E[U(\beta)^2]^{-1/2})\right)^{q_2}\right],
%\E\left[\left(\left|D_ZZW\right|_{\infty}/B_n\right)^{q_2}\right]\right)\le 2$;
%\end{enumerate}
%where $D_Z$ is the population counterpart of $\mathbf{D_Z}$ (i.e., replacing $\mathbb{E}_n$ by $\E$) and $W$ is standard normal random variable independent of $Z$.\\ 

\subsection{Sparsity}

%ALSO TALK ABOUT $d_Z<d_X$: two situations in which sparsity may be used.

In the high-dimensional setting with $d_X$ larger than $n$, a natural and commonly used restriction is that $\beta$ is \textit{sparse}, meaning that it has many elements exactly equal to zero but the researcher does not know which ones. Sparsity implies that there exists an underlying parsimonious model which is unknown to the researcher. It can be used to motivate $\ell_1$ penalized estimators such as the LASSO of \cite{tibshirani96} for regression or the STIV of \cite{gautier21} for instrumental variables.  In the instrumental variables context, sparsity can be interpreted as imposing exclusion restrictions of unknown locations. As explained below, this is particularly useful in the underidentified case with $d_Z<d_X$, which can arise, for example, when there is uncertainty as to which candidate instruments can be excluded, implying that some instruments may be invalid \citep{kolesar15,kang16}. %Hence sparsity is of use for both identification and for high-dimensional estimation. 

The SNIV confidence set can easily accommodate sparsity. We define $S_Q\subseteq [d_X]$ as the indices of the regressors of questionable relevance (i.e., whose entry of $\beta^*$ may be zero). We denote by $d_Q\triangleq|S_Q|$ and modify $\mathcal{I}$ and $\widehat{\mathcal{C}}$ to include the restriction
\begin{align}
	|S(\beta)\cap S_Q|\leq s,\label{sc}
\end{align}
where $s\in[d_Q]$ is an upper bound chosen by the researcher, and recalling that $S(\beta)\subseteq[d_X]$ is the support of $\beta$. Though we do not make it explicit, both $s$ and $S_Q$ can depend on $n$. Since the choice of $s$ provides a guarantee on the sparsity, we refer to it as a \textit{sparsity certificate}. When the sparsity certificate $s$ is used, we use the notation $\mathcal{I}_s$ and $\widehat{\mathcal{C}}_s$ in place of $\mathcal{I}$ and $\widehat{\mathcal{C}}$. Clearly, $\mathcal{I}_{d_Q}=\mathcal{I}$ and $\widehat{\mathcal{C}}_{d_Q}=\widehat{\mathcal{C}}$. %For classes with asymptotic coverage guarantee, both $d_Q$ and $s$ can depend on $n$.

\begin{remark}
The restriction in \eqref{sc} is weaker than imposing $d_Q-s$ exclusion restrictions on the parameters in $S_Q$.
\end{remark}

\begin{remark}
In practice, the researcher may not know how to choose the sparsity certificate. In this case nested confidence sets can be computed by varying $s$ over reasonable alternatives. This allows for an assessment of the information content of progressively stronger assumptions on the sparsity.
\end{remark}

\noindent Interestingly, $\mathcal{I}_s$ can be a singleton even when $\mathcal{I}$ is not. This means that sparsity can lead to point identification even in `underidentified' models (i.e., when $d_Z<d_X$). In general, %and $s<d_Z$, %(i.e., $s<{d_Z}+d_Q-{d_X}$), 
 $\mathcal{I}_s$ is a singleton if there is a solution for only one of the 
${d_Q \choose s}$ overdetermined systems based on  \eqref{einstr}-\eqref{econstraints} 
 and it is unique. For example, $\mathcal{I}_s$ can be a singleton when $s+d_X-d_Q<d_Z<d_X$ and sparsity implies that some %sufficiently many  of the 
exogenous regressors have a zero coefficient (i.e., they are excluded, see \cite{kang16}). The basic idea is that excluded exogenous regressors can serve as instruments for included endogenous regressors, but we need not necessarily know \textit{which} regressors are excluded. %In social effects models, %with observed network 
%others' exogenous characteristics serve as instruments (see \cite{bram}). This paper is applied %in this context 
%to %(partially) 
%unobserved networks in \cite{Rose,GR}, and \cite{BGMPR}.
Finally, if $S_Q=[d_X]$, $\mathcal{I}_s$ is a singleton %it can also be shown that 
%$\mathcal{I}_s$ is a singleton if 
if all matrices formed from $2s$ columns of $\E[ZX^\top]$ have rank $2s$ \citep{candes07}. The corresponding order condition is $s\leq d_Z/2$, which does not depend on $d_X$. 

\begin{figure}[t!]
\centering
\caption{The Identified Set, SNIV Confidence Set and its Outer Approximations}\label{fig1}
\includegraphics[width=0.75\textwidth]{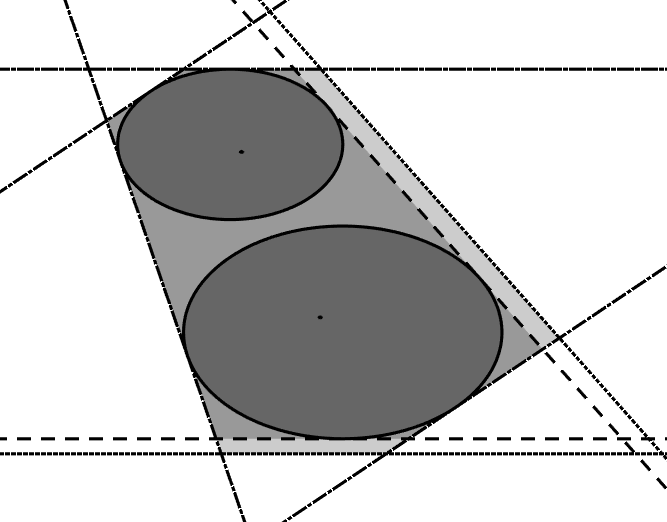}
\begin{tabular}{cccc}
$\mathcal{I}_1$&\includegraphics[width=0.05\textwidth]{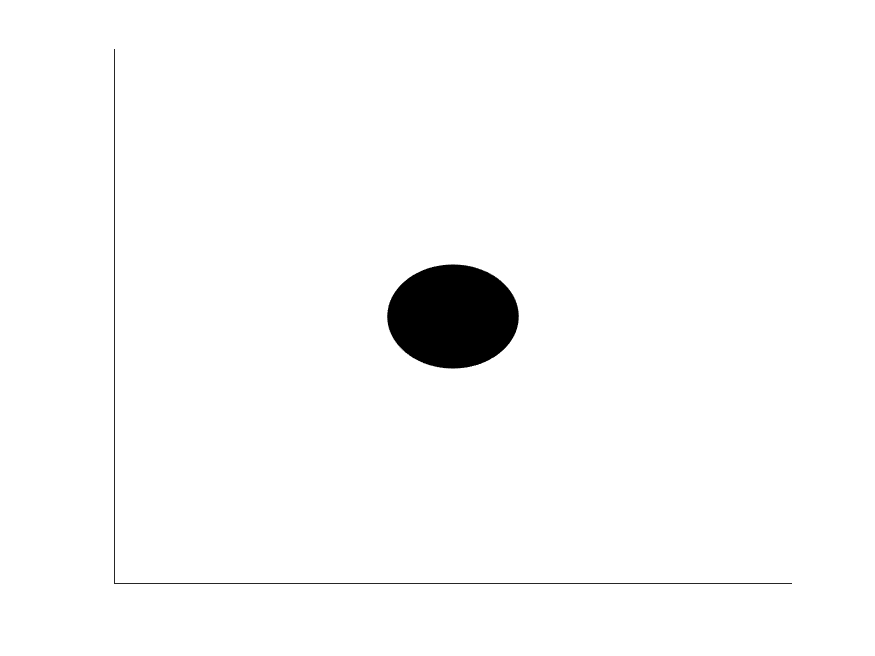}&$\widehat{\mathcal{C}}_1$&\includegraphics[width=0.05\textwidth]{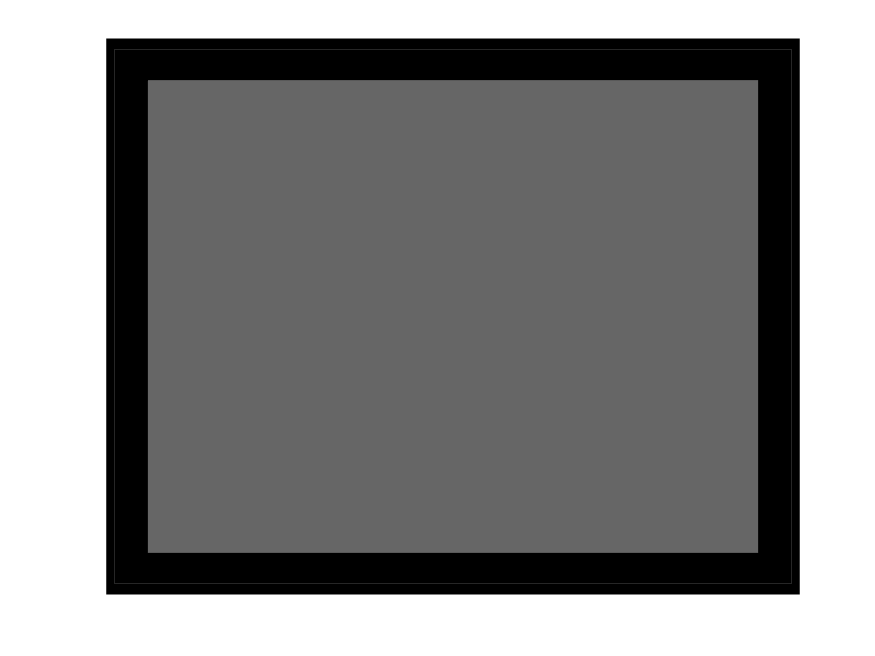}\\
$f_\mathbf{u}(\beta)\geq f^*(\mathbf{u})$&\includegraphics[width=0.05\textwidth]{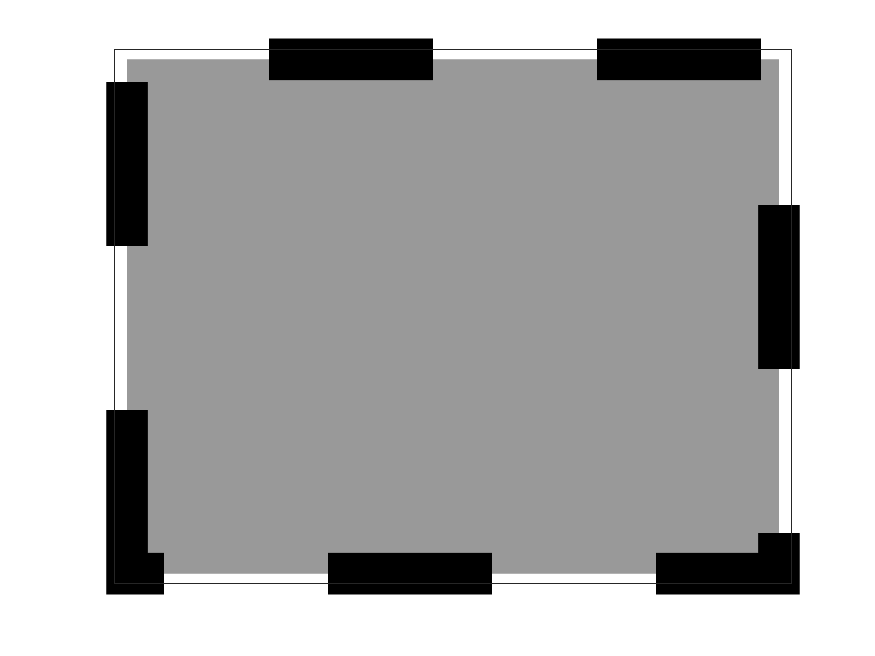}&$f_\mathbf{u}(\beta)\geq f^*_h(\mathbf{u})$&\includegraphics[width=0.05\textwidth]{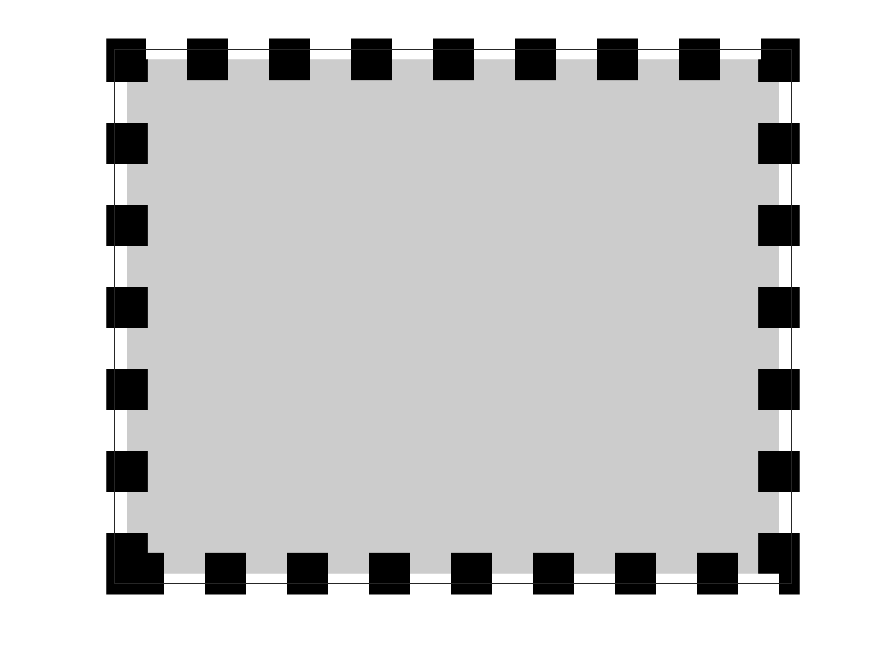}
\end{tabular}\\
\vspace{0.5cm}
\textbf{Notes:} In this example, $d_Z=1$, $d_X=2$, $\mathcal{B}=\R^{d_X}$, $d_Q=[d_X]$ and $s=1$.
\end{figure}

However, SNIV does not require that $\mathcal{I}_s$ be a singleton. For example, $\mathcal{I}_s$ can comprise a finite union of singletons. Figure \ref{fig1} depicts such an example with $d_Z=1$, $d_X=2$, $\mathcal{B}=\R^{d_X}$, $d_Q=[d_X]$ and $s=1$, in which case $\mathcal{I}_s$ is the intersection of the line $\E[Zy]=\E[ZX^\top]\beta$ with the set $\{\beta\in\R^2:\beta_1=0\text{ or } \beta_2=0\}$. The SNIV confidence set allows for such partially identified cases due to the uniformity over $s$ and $\mathcal{I}_s$ in the coverage guarantee, which is obtained by replacing $\inf_{(\beta,\mathbb{P}):\beta\in\mathcal{I}}\mathbb{P}(\beta\in\widehat{\mathcal{C}})$ by $\min_{s\in[d_Q]}\inf_{(\beta,\mathbb{P}):\beta\in\mathcal{I}_s}\mathbb{P}(\beta\in\widehat{\mathcal{C}}_s)$ in Propositions \ref{propcover} and \ref{propcovermboot}.

\subsection{Endogenous instruments}

Testing instrument exogeneity is a classical problem to which our framework can be applied.
%We now consider a high-dimensional framework for the problem of  
%%Testing 
%IV exogeneity (see, e.g., %is a classical problem studied since \cite{ %Bas,
%\cite{sargan58}). %\footnote{This question is addressed in \hyperlink{versions15}{(v1)}, see \hyperlink{versions15}{(v5)} for references.}. % for the linear IV model. % and in \cite{Hansen} for GMM (see also \cite{And,CL}). %\cite{Liao}, 
%%Cheng and Liao (2015)). 
Introducing $\theta\in\R^{d_Z}$ to account for the possible failure of exogeneity,  
%In the 
we replace \eqref{einstr}-\eqref{econstraints} by
\begin{align}
&\E[ZU(\beta)-\theta]=0,\label{einstrNV}\\
&\left(\beta,\theta\right)\in\mathcal{B}\times\Theta,\ \mathbb{P}\left(\beta,\theta\right)\in\mathcal{P},\label{econstraintsNV}
\end{align}
where $\theta_l\ne0$ means that $Z_l$ is endogenous, $\mathbb{P}\left(b,t\right)$ is the distribution of $\left(X,Z,ZU(b)-t\right)$ implied by $\mathbb{P}$ and $\Theta\subseteq\R^{d_Z}$ encodes restrictions on $\theta$. %The class $\mathcal{P}$ can be any of classes 1-3, replacing $\bold{Z}_{i,\cdot}^\top\bold{U}_i(\beta)$ with $\bold{Z}_{i,\cdot}^\top\bold{U}_i(\beta)-\theta$.
For example, $\Theta$ may be such that the sign of the correlation of a regressor and the structural error is known. %or that she has imperfect instruments %which have smaller correlation with the structural error than the endogenous regressor 
%(see \cite{nevo}).
An important restriction encoded by $\Theta$ is $\theta_{ S_{\perp}}=0$ for $S_{\perp}\subseteq[d_Z]$, which indexes the instruments known to be exogenous. The remaining instruments are potentially endogenous. We can use a sparsity certificate to place an upper bound on the number of endogenous instruments, given by
\begin{align}
|S(\theta)|\le \widetilde{s},\label{sct}
\end{align}
for a given $\widetilde{s}\in[\widetilde{d}_Q]$, where $\widetilde{d}_Q\triangleq d_Z-|S_\perp|$. Thus, though the identities of the endogenous instruments may not be known, their number can be restricted. The counterpart of $\mathcal{I}_s$, denoted by $\mathcal{I}_{s,\widetilde{s}}$, collects the vectors which satisfy \eqref{einstrNV}-\eqref{econstraintsNV} and the sparsity restrictions in \eqref{sc} and \eqref{sct}. Under Classes 1-3,\footnote{Class 4 is not applicable with possibly endogenous instruments.} SNIV is
\begin{align}
	\widehat{\mathcal{C}}_{s,\widetilde{s}}\triangleq\left\{(\beta,\theta)\in\mathcal{B}\times\Theta:\begin{array}{l} |\mathbf{D}(\beta,\theta)(\E_n[ZU(\beta)]-\theta)|_\infty\leq r_n,\\|S(\beta)\cap S_Q|\leq s,|S(\theta)|\le \widetilde{s}\end{array} \right\},
\end{align}
where $\mathbf{D}(\beta,\theta)$ is the $d_Z\times d_Z$ positive, diagonal matrix with $l^{\text{th}}$ diagonal element $\mathbb{E}_n[(Z_lU(\beta)-\theta_l)^2]^{-1/2}$. This set allows one to simutaneously perform inference on $\beta^*$ and $\theta^*$, without requiring, for example, a pilot estimator and subsequent test of instrument exogeneity. The coverage guarantee is obtained by replacing $\inf_{(\beta,\mathbb{P}):\beta\in\mathcal{I}}\mathbb{P}(\beta\in\widehat{\mathcal{C}})$ by $\min_{\widetilde{s}\in [\widetilde{d}_Q]}\min_{s\in[d_Q]}\inf_{(\beta,\theta,\mathbb{P}):(\beta,\theta)\in\mathcal{I}_{s,\widetilde{s}}}\mathbb{P}((\beta,\theta)\in\widehat{\mathcal{C}}_{s,\widetilde{s}})$ in Proposition \ref{propcover}. Both $\widetilde{S}_Q$ and $\widetilde{s}$ can depend on $n$. 

%In sum, the SNIV confidence set can allow for many, weak and endogenous instruments as well as uncertain exclusion restrictions, is robust to identification, does not restrict the number of regressors and, is uniform and, under Classes 2-3, delivers a finite sample coverage guarantee.

%The SNIV confidence set is very general. It  allows for conditional heteroskedasticity and requires no assumption on the relative magnitudes of $n,d_X$ and $d_Z$ nor on the joint distribution of $X$ and $Z$. Hence, SNIV is robust to identification (including partial identification) and can simultaneously allow for many, few, weak and endogenous instruments, as well as uncertain exclusion restrictions. SNIV easily accommodates the researcher's knowledge on parameter restrictions ($\mathcal{B},\Theta$), and sparsity ($S_Q,s,S_\perp,\widetilde{s}$), and has good statistical properties regarding uniformity and finite sample validity. We now provide a computationally tractable implementation.

\section{Computation}\label{comp}

To implement SNIV the researcher needs some way to summarize the vectors which lie in the confidence set. \cite{belloni12} propose to use a grid for a confidence set with no sparsity constraints nor potentially endogenous instruments. This involves checking whether the inequalities in the definition of the SNIV confidence set are verified for every $\beta$ on a grid over $\mathcal{B}$, and is a practical solution when $d_X$ is small. However, a grid search quickly becomes infeasible for moderate $d_X$. In a low-dimensional setting, one can first partial-out a small number of exogenous regressors (see Remark \ref{r1}) so that $d_X$ is the number of endogenous regressors, which may be sufficiently small so as to use a grid. Otherwise we require an alternative.

We propose a method based on solving convex optimization problems. %Defining $\gamma\triangleq(\beta^\top,\theta^\top)^\top\in\Gamma \triangleq\mathcal{B}\times\Theta$, 
For a given direction $\mathbf{u}\in\R^{d_X+d_Z}$ normalized to satisfy $|\mathbf{u}|_2=1$ and the function $f_{\mathbf{u}}(\beta,\theta)\triangleq\mathbf{u}^\top(\beta^\top,\theta^\top)^\top$ we seek to compute
\begin{align}
f^*(\mathbf{u})\triangleq \inf_{(\beta,\theta)\in\widehat{\mathcal{C}}_{s,\widetilde{s}}}f_\mathbf{u}(\beta,\theta),\label{prob}
\end{align}
which is the support function of $\widehat{\mathcal{C}}_{s,\widetilde{s}}$. By solving \eqref{prob} for all directions $\mathbf{u}\in\{\mathbf{u}\in\R^{d_X+d_Z}:|\mathbf{u}|_2=1\}$, we obtain the convex envelope of $\widehat{\mathcal{C}}_{s,\widetilde{s}}$ defined by the inequalities $f_\mathbf{u}(\beta,\theta)\geq f^*(\mathbf{u})$ for all $\mathbf{u}\in\{\mathbf{u}\in\R^{d_X+d_Z}:|\mathbf{u}|_2=1\}$.\footnote{In practice, we  consider only a finite number of directions.}% $\mathcal{U}\subset\{\mathbf{u}\in\R^{d_X+d_Z}:|\mathbf{u}|_2=1\}$.

If $\widehat{\mathcal{C}}_{s,\widetilde{s}}$ is convex, solving \eqref{prob} is straightforward. In general $\widehat{\mathcal{C}}_{s,\widetilde{s}}$ is not convex because none of the inequalities in its definition define a convex set. This is unavoidable because $\mathcal{I}_{s,\widetilde{s}}$ need not be convex. We now show that $\widehat{\mathcal{C}}_{s,\widetilde{s}}$ is a semi-algebraic set (i.e., a set defined by polynomial inequalities) and apply methods in semi-algebraic optimization to problem \eqref{prob}.
\begin{proposition}\label{propsnivg}
If $\mathcal{B}$ and $\Theta$ are semi-algebraic, the SNIV confidence set is semi-algebraic, taking the form
\begin{align}
\widehat{\mathcal{C}}_{s,\widetilde{s}}=\left\{(\beta,\theta): \exists\gamma\in[0,1]^{d_Q+\widetilde{d}_Q}: \widehat{\mathbf{g}}(\beta,\theta,\gamma)\geq 0 \right\}
\end{align}
where $\widehat{\mathbf{g}}$ is a $d_g\times 1$ vector of polynomials, the form of which is given in the proof. If $\mathcal{B}$ and $\Theta$ are defined by polynomial inequalities of degree at most $e$ then $\widehat{\mathbf{g}}$ has degree $\max(2,e)$.
\end{proposition}
The requirement that $\mathcal{B}$ and $\Theta$ are semi-algebraic is mild. For example, $\mathcal{B}=\mathbb{R}^{d_x}$ is semi-algebraic. The additional parameter $\gamma$ is required to model the sparsity constraints in \eqref{sc} and \eqref{sct}. Proposition \ref{propsnivg} implies that
\begin{align}
f^*(\mathbf{u})=\inf_{(\beta,\theta,\gamma): \widehat{\mathbf{g}}(\beta,\theta,\gamma)\geq 0}f_{\mathbf{u}}(\beta,\theta),\label{prob1}
\end{align}
which can be computed by solving a polynomial optimization problem. Due to non-convexity, exact computation of $f^*(\mathbf{u})$ is NP-hard. Instead, we focus on solving convex relaxations of \eqref{prob1}. Convex relaxation is routinely used to construct computationally tractable estimators. For example, LASSO uses an $\ell_1$ penalty as a convex relaxation of a sparsity constraint such as \eqref{sc}. We solve a sequence of convex relaxations, delivering a \textit{hierarchy} of convex optimization problems. Following the seminal paper of \cite{lasserre01}, such hierarchies have attracted much attention in the optimization literature in recent years. We first provide a general summary of the approach, then explain the specific hierarchy we propose.

The most important feature of a hierarchy is that it is \textit{disciplined}, meaning that it delivers a monotone sequence of lower bounds converging to $f^*(\mathbf{u})$. If $f^*_h(\mathbf{u})$ is the optimal value obtained by solving the $h^{\text{th}}$ convex optimization problem in the hierarchy, we have $f^*_{h}(\mathbf{u})\leq f_{h+1}^*(\mathbf{u})\leq f^*(\mathbf{u})$ for all $h\in\mathbb{N}$ and $f^*_h(\mathbf{u})\to f^*(\mathbf{u})$ as $h\to\infty$. As $h$ increases, though convex, the optimization problems become more computationally intensive. It is also generically the case that there exists finite $h^*$ such that $f^*_{h^*}(\mathbf{u})=f^*(\mathbf{u})$, and that the researcher can identify when such $h^*$ has been encountered. %Handbook-style expositions can be found in \cite{anjos11} and \cite{lasserre15}. 

Monotonicity of the sequence of lower bounds on $f^*(\mathbf{u})$ is crucial. This is because it allows us to construct bounds on the convex envelope of the SNIV confidence set defined by the linear inequalities $f_\mathbf{u}(\beta,\theta)\geq f^*_{h}(\mathbf{u})$ for all $\mathbf{u}\in\mathcal{U}$ and some $h\in\mathbb{N}$, where $\mathcal{U}$ is a finite collection of directions. Since we construct a superset, the coverage guarantee cannot fall below $1-\alpha$. The larger is $h$, the closer the superset becomes to the SNIV confidence set. Hence, by varying $h$ and $\mathcal{U}$, we can trade off the computational burden with the quality of the approximation without compromising the coverage guarantee. Such a trade-off cannot be achieved by local or heuristic optimization methods nor by adjusting the spacing of a grid, both of which may compromise the coverage guarantee. Figure \ref{fig1} illustrates the SNIV confidence set and its outer approximations for a partially identified model with $d_Z=1$, $d_X=2$, $\mathcal{B}=\R^{d_X}$, $d_Q=[d_X]$, $s=1$ and all instruments known to be exogenous.

\subsection{Low-dimensional objects of interest}\label{remci}

The method we propose can also be used to compute bounds on a polynomial function of interest $p(\beta^*,\theta^*)$. For example, to obtain a confidence interval for $\beta_1^*$, we can solve \eqref{prob} for $f_{\mathbf{u}_1}$ and $f_{\mathbf{u}_2}^\top$, where $\mathbf{u}_1=(1,0,...,0)^\top$ and $\mathbf{u}_2=-\mathbf{u}_1$ (i.e., use the projection method). More generally one can consider a vector $\mathbf{p}$ of functions of interest. If the dimension of $\mathbf{p}$ is small relative to $d_X$, it is well known that the projection method can be conservative. A leading case with $d_X=\dim(\mathbf{p})=1$ is a confidence interval in a model with one endogenous regressor. In this case, the projection method is not conservative and SNIV is simultaneously robust to weak instruments and to $d_Z$ much larger than $n$.

\subsection{A hierarchy of semidefinite optimization problems}

Since it is straightforward to implement, %\footnote{We use Gloptipoly 3, available at \url{https://homepages.laas.fr/henrion/software/gloptipoly3/}.}
we present our application of the seminal hierarchy first proposed by \cite{lasserre01}. This hierarchy is sufficiently computationally tractable to deal with problems of size likely to be encountered in empirical work. Recent advances allowing for even larger problems are provided by \cite{lasserre17} and \cite{weisser18}. 

To simplify the exposition, we denote the decision variable in problem \eqref{prob1} by $\delta\triangleq(\beta^\top,\theta^\top,\gamma^\top)^\top$ of size $d_\delta\triangleq d_X+d_Z+d_Q+\widetilde{d}_Q$. %We denote a monomial of $\mu$ by $\mu_{\mathbf{a}}$, where $\mathbf{a}$ is a vector of exponents for each entry of $\mu$. For example, for the monomial $\mu_1^2$ we use $\mathbf{a}=(2,0,0...,0)$. 
The hierarchy uses the decision variable $\mu$, each entry of which represents a monomial of $\delta$. For example, if $d_\delta=2$ then $\mu=(1,\delta_1,\delta_2,\delta_1^2,\delta_1\delta_2,\delta_2^2,...)^\top$, so the polynomial $p(\delta)=\delta_2+2\delta_1^2$ is equivalently expressed as $\mu_3+2\mu_4$. This allows us to define the \textit{Riesz linear functional} of $p$ as $L_\mu(p)=\mu_3+2\mu_4$.\footnote{We present the case in which he order of the entries of $\mu$ is a graded lexicographic order. Other orderings are possible. None of our results depend on the ordering used.} Now let the vector $\mathbf{m}_e(\delta)$ comprise all monomials of $\delta$ of degree no larger than $e$. For example, $\mathbf{m}_1(\delta)=(1,\delta_1,\delta_2)^\top$. Then we can define the \textit{moment matrix} $\mathbf{M}_e(\mu)\triangleq L_\mu(\mathbf{m}_e(\delta)\mathbf{m}_e(\delta)^\top)$. For example, if $d_\delta=2$ and $e=1$, we have
\begin{align}
\mathbf{m}_1(\delta)\mathbf{m}_1(\delta)^\top=\begin{bmatrix}1&\delta_1&\delta_2\\ \delta_1&\delta_1^2&\delta_1\delta_2\\\delta_2&\delta_1\delta_2&\delta_2^2\end{bmatrix}\Rightarrow \mathbf{M}_1(\mu)=\begin{bmatrix}\mu_1&\mu_2&\mu_3\\ \mu_2&\mu_4&\mu_5\\\mu_3&\mu_5&\mu_6\end{bmatrix}.
\end{align}
Given another polynomial $q(\delta)$, we can similarly define the \textit{localizing matrix} $\mathbf{M}_e(q \mu)\triangleq L_\mu(q(\delta)\mathbf{m}_e(\delta)\mathbf{m}_e(\delta)^\top)$.

%Given the above definitions, we can now define the hierarchy for problem \eqref{prob1}. 
At level $h$ of the hierachy we solve the semidefinite program
\begin{align}
f^*_h(\mathbf{u})\triangleq\inf_{\mu}L_\mu(f_\mathbf{u})\quad\text{subject to}\quad&\mathbf{M}_h(\mu)\succcurlyeq \mathbf{0},\nonumber\\
&\mathbf{M}_{h-e_j}(\widehat{\mathbf{g}}_j\mu)\succcurlyeq \mathbf{0},\quad j\in[d_g],\nonumber\\
&\mu_1=1,\label{sdp}
\end{align}
where $e_j$ is the smallest integer which is at least as large as deg$(\widehat{\mathbf{g}}_j)/2$ for $j\in[d_g]$. This program has a linear objective function and $d_g+1$ semidefinite constraints. The semidefinite constraint on $\mathbf{M}_h(\mu)$ arises because $\mathbf{m}_h(\delta)\mathbf{m}_h(\delta)^\top$ has rank 1. In principle we would like to impose that $\mathbf{M}_h(\mu)$ has rank 1. However, the set of rank 1 matrices is not convex. To obtain a convex problem, we use instead the set of positive semidefinite matrices. The intuition is the same for the other $d_g$ semidefinite constraints because the polynomials $\widehat{\mathbf{g}}$ are restricted to be non-negative.
\begin{corollary}\label{corconverge}
If $\mathcal{B}$ and $\Theta$ are compact then $f^*_{h}(\mathbf{u})\leq f_{h+1}^*(\mathbf{u})\leq f^*(\mathbf{u})$ for all $h\in\mathbb{N}$ and $f^*_h(\mathbf{u})\to f^*(\mathbf{u})$ as $h\to\infty$.
\end{corollary}
Corollary \ref{corconverge} follows from Proposition \ref{propsnivg} due to Theorem 4.2 of \cite{lasserre01}. The only assumption beyond the class $\mathcal{P}$ is a technical assumption requiring that the parameter space be compact. Compactness is useful because it allows us to find $B$ sufficiently large such that the redundant polynomial constraint 
\begin{align}
B-|\beta|_2^2-|\theta|_2^2\geq 0\label{B}
\end{align} 
holds. In practice, we augment the constraints $\widehat{\mathbf{g}}(\beta,\theta,\gamma)\geq 0$ to include \eqref{B} prior to applying the semidefinite hierarchy. %This is a simple way to guarantee that the high-level assumptions of Theorem 4.2 of \cite{lasserre01} are satisfied \citep{lasserre01}. 

Though compactness is a common technical assumption, in practice we may often not have compact $\mathcal{B}$ and $\Theta$. For example, we may have $\mathcal{B}=\R^{d_X}$. In this case we suggest increasing $B$ until \eqref{B} ceases to bind at the solution. If the SNIV confidence set is unbounded in direction $\mathbf{u}$, \eqref{B} will always bind. In practice this is of little consequence since there is little distinction between $f^*(\mathbf{u})$ being $-\infty$ or an arbitrarily small finite constant. Thus, when the parameter space is not compact, our approach characterizes the intersection of the SNIV confidence set with an arbitrarily large $\ell_2$ ball.

The intuition for the result that $f^*_{h}(\mathbf{u}) \leq f^*(\mathbf{u})$ for all $h\in\mathbb{N}$ comes from convex relaxation. By replacing rank 1 constraints for the moment and localizing matrices by positive semidefinite constraints, we minimize over a larger set, hence it must be that we obtain a lower bound on the optimal value. The intuition for $f^*_{h}(\mathbf{u}) \leq f_{h+1}(\mathbf{u})$ for all $h\in\mathbb{N}$ is that increasing $h$ reduces the size of the set over which we minimize, hence must always deliver a larger optimal value. The computational trade-off is also clear from the form of problem \eqref{sdp} because the dimension of the moment matrix $\mathbf{M}_h(\mu)$ is ${d_\delta+h \choose h}$, which is increasing in $h$. Similarly, the dimensions of the localizing matrices are combinatorically increasing in $h$. Thus, increasing $h$ delivers a tigher bound but at increased computational cost.

We implement the hierarchy using the following algorithm proposed by \cite{lasserre15}.
\begin{algorithm}
Inititialize $h=1$ and the largest level of the hierachy $\overline{h}\in\mathbb{N}$. Then,
	\begin{enumerate}
		\item Solve the semidefinite optimization problem in \eqref{sdp} to obtain optimal value $f^*_h(\mathbf{u})$ and optimal solution $\mu^*$ (if it exists).
		\item If there is no optimal solution $\mu^*$ and $h<\overline{h}$ then increase $h$ by one and go to step 1.  If there is no optimal solution $\mu^*$ and $h=\overline{h}$ then terminate the algorithm with bound $f^*_{h-1}(\mathbf{u})$.
		\item If $\text{rank }\mathbf{M}_h(\mu^*)=\text{rank }\mathbf{M}_{h-e}(\mu^*)$ (where $e\triangleq \max_{j\in[d_g]}e_j$) then we know $f^*_h(\mathbf{u})=f^*(\mathbf{u})$. Terminate the algorithm with the exact bound $f^*_h(\mathbf{u})$.
		\item  If $\text{rank }\mathbf{M}_h(\mu^*)\neq \text{rank }\mathbf{M}_{h-e}(\mu^*)$ and $h<\overline{h}$ then increase $h$ by one and go to step 1. Else, if $h=\overline{h}$, terminate the algorithm with the lower bound $f^*_h(\mathbf{u})$.
	\end{enumerate}
\end{algorithm}
The basic idea is to begin with the most computationally tractable semidefinite program with $h=1$ and continue to increase $h$ until either we know that $f^*_h(\mathbf{u})=f^*(\mathbf{u})$ (step 3) or we hit the largest computationally feasible level of the hierarchy ($\overline{h}$). In practice, $\overline{h}$ is determined by the size of the problem and the available computational resources. In our Monte-Carlo experiment we use $\overline{h}=2$ on a standard desktop machine. Step 3 provides a stopping criterion which can be used to establish finite convergence of the hierarchy. For brevity, we do not provide technical conditions under which finite convergence is possible, which can be found in \cite{lasserre15} (see Theorem 6.5) and involve standard Karusch-Kuhn-Tucker conditions for an optimal solution to be a local minimizer of a nonlinear program. In fact, these conditions imply that finite convergence is achieved generically \citep{lasserre15} (see Theorem 7.6), though there is no guarantee that it is achieved for small values of $h$. In our Monte-Carlo experiment we achieve finite convergence with high frequency in some designs but with low frequency in others.%In the special case in which the SNIV confidence set is convex, finite convergence is guaranteed \citep{lasserre09}.

\section{Inverting other robust tests}\label{classic}

%Though we focus the exposition on linear instrumental variables models, the approach can be used to invert any test for which the non-rejection region is semi-algebraic, providing a computationally attractive alternative to a grid search. We briefly discuss partially identified models in Section \ref{partial}.
In the standard linear instrumental variables setting with large $n$ and fixed $d_Z\geq d_X$, robust inference can be conducted by inverting robust tests \citep{andrews19}. In this section we show that the computational approach of Section \ref{comp} can be applied to do so. There are myriad such tests, including but not limited to, the Anderson Rubin (AR) test \citep{anderson49}, Lagrange-multiplier (LM) test \citep{kleibergen02,moreira02} and the Conditional Likelihood Ratio (CLR) test \citep{moreira03}. All of these tests have a non-rejection region (i.e., a confidence set) of the form
\begin{align}
	\widetilde{\mathcal{C}}\triangleq\{\beta\in\mathcal{B}: \widehat{p}(\beta)\leq \widehat q_\alpha(\beta) \},\label{gineq}
\end{align}
where $\widehat{p}$ and $\widehat q_\alpha$ are polynomials and the coefficients of $\widehat q_\alpha$ depend on the confidence level $\alpha$. As with the SNIV confidence set, $\widetilde{\mathcal{C}}$ is semi-algebraic whenever $\mathcal{B}$ is. The polynomial inequality in the definition of $\widetilde{\mathcal{C}}$ can be degree 2 (AR test, CLR test with $d_X=1$) or larger (LM test, CLR test with $d_X>1$). For example, the AR test under homoskedasticity uses $\widehat{p}(\beta)=\widehat{p}_{AR}(\beta)$ and $\widehat q_\alpha(\beta)= \widehat q_{AR,\alpha}(\beta)$, where
\begin{align}
\widehat p_{AR}(\beta)&\triangleq\mathbf{U}(\beta)^\top\mathbf{Z}(\mathbf{Z}^\top\mathbf{Z})^{-1}\mathbf{Z}^\top\mathbf{U}(\beta),\quad\widehat q_{AR,\alpha}(\beta)\triangleq C_{\alpha}(d_Z) \widehat{q}_{AR}(\beta), \label{ar1}\\
\widehat{q}_{AR}(\beta)&\triangleq(1,-\beta^\top)(\mathbf{y},\mathbf{X})^\top(\mathbf{I}_n-\mathbf{Z}(\mathbf{Z}^\top\mathbf{Z})^{-1}\mathbf{Z}^\top)(\mathbf{y},\mathbf{X})(1,-\beta^\top)^\top/(n-d_Z),\label{ar2}
\end{align}
$\mathbf{y}$ is the $n\times 1$ vector of outcomes, $\mathbf{X}$ is the $n\times d_X$ matrix of regressors, $\mathbf{U}(\beta)\triangleq\mathbf{y}-\mathbf{X}\beta$, $\mathbf{Z}$ is the $n\times d_Z$ matrix of instruments and $C_{\alpha}(d)$ is the $1-\alpha$ quantile of the $\chi^2_{d}$ distribution. None of the above tests yield a convex confidence set \citep{mikusheva10}, making a grid search computationally demanding (see \cite{andrews16con}, supplementary material). Alternatives to a grid search (e.g., \cite{mikusheva10} for the CLR test) can also be computationally intensive for moderate $d_X$. Hierarchies of semidefinite optimization problems provide a practical alternative.

Sometimes the object of interest may be a function of $\beta^*$ of dimension smaller than $d_X$ (see Section \ref{remci}). For example, one may be interested in a sub-vector of $\beta^*$. To obtain coverage probability $1-\alpha$ for a confidence set for a sub-vector (e.g., a confidence interval), we need to adjust $\widehat{q}_\alpha$. \cite{guggenberger12}, \cite{guggenberger19} and \cite{guggenberger21} provide appropriate adjustments for the AR test. Decomposing $\beta=(\beta_1^\top,\beta_2^\top)^\top$, the results of \cite{guggenberger12} imply that an asymptotic $1-\alpha$ confidence set for $\beta_1^*$ under homoskedasticity is
\begin{align}
	\widetilde{\mathcal{C}}_{AR}\triangleq\left\{\beta_1:\inf_{\beta_2} \frac{\widehat p_{AR}(\beta)}{\widehat{q}_{AR}(\beta)}\leq C_{\alpha}(d_Z-d_X+d_{X_1})  \right\},\label{gineqAR}
\end{align}
where $d_{X_1}$ is the dimension of $\beta_1$.\footnote{This decomposition is without loss of generality because the regressors can be reordered.} Thus, for a given direction $\mathbf{u}_1\in\R^{d_{X_1}}$ normalized to have $|\mathbf{u}_1|_2=1$ we seek to compute $\inf_{\beta_1\in\widetilde{\mathcal{C}}_{AR}}\mathbf{u}_1^\top\beta_1$, equivalently expressed as
\begin{align}
\inf_{\substack{\beta: \widehat p_{AR}(\beta)\leq  C_{\alpha}(d_Z-d_X+d_{X_1}) \widehat{q}_{AR}(\beta)}}\mathbf{u}^\top\beta,\label{probAR}
\end{align}
where $\mathbf{u}=(\mathbf{u}^\top_1,\mathbf{0}^\top)^\top$ is $d_X\times 1$. This is a polynomial optimization problem, hence we can apply convex hierarchies to find a monotonic sequence of lower bounds. For the special case of a confidence interval we have $d_{X_1}=1$ hence only need consider $\mathbf{u}_1=\pm 1$. \cite{guggenberger19} replace $ C_{\alpha}(d_Z-d_X+d_{X_1})$ with an alternative which delivers a less conservative confidence set in a finite sample and \cite{guggenberger21} extend the approach to allow for conditional heterokskedasticity. Thus, we can apply hierarchies of semidefinite optimization problems to \eqref{probAR} in order to obtain AR confidence intervals which can be rapidly computed.

\section{Monte-Carlo}\label{monte}

To illustrate our approach, we consider a setting with $d_X=10$ endogenous regressors. We choose this design because $d_X$ is large enough to render a grid search infeasible yet small enough to permit many replications of our experiment on a standard desktop machine within a reasonable timeframe.%\footnote{We use a machine with 32GB of RAM and a 3.6GhZ quad core processor.} 

We consider an i.i.d. sample of size $n=2000$ satisfying \eqref{einstr}-\eqref{econstraints}. The instruments are related to the regressors according to $\mathbb{E}[ZV(\Pi)]=0$ where $V(\Pi)\triangleq X-\Pi Z$ and $\Pi$ is $d_X\times d_Z$. We set $\beta^*=(1,-1,0,...,0)^\top$ and vary $\Pi^*$ by design, as explained below. The instruments follow $\mathcal{N}(0,I_{d_Z})$ and the error terms verify $(U(\beta^*),V(\Pi^*)^\top)^\top\sim\mathcal{N}(0,\Omega)$, where $\Omega_{11}=1$ (homoskedasticity) or $\Omega_{11}=Z_1^2$ (conditional heteroskedasticity), $\Omega_{1j}=(-1)^j(1-\pi^{*})/5$, $\Omega_{jj}=1-\pi^{*}$ for $j>1$ and all other entries are equal to zero. The parameter $\pi^*\in[0,1]$ determines the fraction of the variance of each regressor which is due to the instruments. In all designs, the variances of each regressor and the structural error are equal to 1. 

To compute the SNIV confidence set we choose $r_n$ using Class 1 and Class 3 with $\alpha=0.05$. To implement the hierarchy, we use $\overline{h}=2$ and $B=1000$. We compute the coverage probability for the SNIV confidence set, and, for designs in which it is feasible, the AR confidence set and confidence intervals. For the SNIV and AR confidence sets, we also report the coverage probability for their outer approximations obtained by solving hierarchies of semidefinite optimization problems, defined by $f_\mathbf{u}(\beta)\geq f^*_h(\mathbf{u})$ for all $\mathbf{u}\in\mathcal{U}$, where, for $\mathcal{U}$ we use a grid of 1600 points over the surface of an $\ell_2$ ball of radius 1.\footnote{In practice we parallelize over $\mathbf{u}\in\mathcal{U}$.} 

Our results are collected in Table \ref{tab:sims}. We focus the discussion of SNIV on Class 1, which, identically to AR, provides an asymptotic coverage guarantee. Class 3 provides a finite guarantee, hence a larger confidence set in all designs. Nevertheless, we find that whenever Class 1 provides an informative confidence set, so does Class 3.\\% Class 3 is also slower to compute, though still sufficiently fast to conduct informative inference using a standard machine. \\

\noindent{\textbf{Classical design.}} We set $d_Z=d_X=10$, $\pi^*=0.3$ and $\Pi^*=\sqrt{\pi^*}I_{d_Z}$ and do not impose any sparsity constraint. The SNIV and AR confidence sets have similar coverage, both of which are marginally below the nominal level. The SNIV confidence set is marginally narrower than the AR confidence set. The coverage of the AR confidence intervals are almost exactly equal to the nominal level, and their width is narrower than either of the confidence sets, as expected. Almost all optimization problems solved yielded an exact global optimum (i.e., $f^*_h(\mathbf{u})=f^*(\mathbf{u})$). The time taken to solve an optimization problem is a little over one second for SNIV and the AR confidence set/interval. In the design with conditional heteroskedasticity the SNIV confidence set is marginally wider than under homoskedasticity and attains the nominal coverage.\\

\noindent{\textbf{Many instruments.}} We take the classical design and add 1989 redundant instruments, all drawn from the standard normal distribution. This yields $d_Z=1999$ instruments and $n=2000$ observations. The SNIV confidence sets have coverage almost identical to the nominal level but are wider than the classical design. However, they remain sufficiently narrow as to be informative on the sign of the nonzero entries of $\beta^*$. In contrast, the AR confidence sets and intervals do not have the correct coverage and are too wide so as to be informative. We also consider an identical design but with $d_Z=2100$. The SNIV confidence sets are similar to the case of $d_Z=1999$, whereas AR confidence sets and intervals are not defined. Almost all optimization problems solved yielded a global optimum. The SNIV optimization problems are solved more slowly than the classical design, taking around 4 seconds on average. In the designs with conditional heteroskedasticity the SNIV confidence set has nominal coverage but is marginally narrower than under homoskedasticity, likely because a greater fraction of the optimization problems yielded an exact global optimum. \\

\noindent{\textbf{Weak instruments.}} We take the classical design and set $\pi^*=0.03$. The AR confidence sets are narrower than SNIV but have coverage further from the nominal level. The AR confidence intervals have coverage slightly larger than the nominal level. Almost all optimization problems solved yielded a global optimum. In the design with conditional heteroskedasticity the SNIV confidence set is marginally wider than under homoskedasticity with coverage close to the nominal level. Computation timings are similar to the classical design.\\

\noindent{\textbf{Invalid instruments.}} We take the classical design with one endogenous regressor and $d_Z=9$ instruments, all of which are included as regressors (i.e., $X_k$=$Z_{k-1}$ for $k=2,3,...,d_X$). Hence there are $d_X=10$ regressors, but only the first is endogenous. We set $\Pi^*=(0,0,...,0,\sqrt{\pi^*}/2,-\sqrt{\pi^*}/2)^\top$ so that only the final two instruments are correlated with the endogenous regressor. We suppose that $S_Q=[d_X]$ (i.e., the relevance of all regressors is questionable) and consider the sparsity certificates $s\in\{2,3\}$ (recalling that $\beta^*$ has two nonzero entries). This design is such that $\mathcal{I}_2$ is a singleton, $\mathcal{I}_3$ is not a singleton but is bounded, and $\mathcal{I}_s$ is unbounded for $s>3$. When $s=3$, though $\beta^*_1=1$ we have $\min_{\beta\in\mathcal{I}_3} \beta_1=0$ and $\max_{\beta\in\mathcal{I}_3} \beta_1=1$. The AR confidence sets and intervals cannot be computed.

The SNIV confidence set has coverage slightly larger than the nominal level. For $s=2$, SNIV is sufficiently narrow so as to be informative on the sign of the nonzero entries of $\beta^*$. For $s=3$, the width of SNIV for $\beta_1$ is around 1.25 on average, which is not sufficiently narrow so as to be informative on the sign of $\beta^*_1$. This is expected because $\beta^*_1$ is not point identified (the identified set has width 1), as explained in the previous paragraph. In the design with conditional heteroskedasticity the SNIV confidence set is marginally wider than under homoskedasticity. Each optimization problem is solved in around 17-27 seconds depending on the sparsity certificate used. This is likely due to the non-convex nature of the sparsity constraint. Nevertheless, the problems are sufficiently tractable so as to allow informative inference on a standard machine.\\

\noindent{\textbf{Endogenous instruments.}} We take the classical design and but add the instruments $Z_{11}=X_1$ and $Z_{12}=X_2$ (i.e., include the first two regressors as additional instruments). This results in $d_Z=12$ instruments, two of which are endogenous, with $\theta_{11}=\Omega_{1,2}$ and $\theta_{12}=\Omega_{1,3}$. We suppose that the researcher questions the exogeneity of the two endogenous instruments and the final three exogenous instruments, hence $S_\perp=[7]$. This implies that are seven instruments known to be exogenous, whereas $d_X=10$, hence the model using only the instruments known to be exogenous is underidentified. Thus, classical tests of overidentifying restrictions are infeasible.

Since the identified set is otherwise unbounded, we restrict the number of endogenous instruments using the sparsity certificate $\widetilde{s}$ on $\theta^*$. We do not make any sparsity restriction on $\beta^*$. Using $\widetilde{s}=2$ corresponds to the case where we assume that there are ten exogenous instruments, but we do not know all of their identities. This design is such that $\mathcal{I}_{d_X,2}$ is a singleton. In contrast, $\mathcal{I}_{d_X,\widetilde{s}}$ is unbounded for $\widetilde{s}>2$. We compute the SNIV confidence set for $\widetilde{s}=2$.

The SNIV confidence set has coverage almost exactly equal to the nominal level and is sufficiently narrow so as to be informative on the sign of the nonzero entries of $\beta^*$. Moreover, the SNIV confidence set allows the null hypotheses of $\theta_{S_\perp^c}=0$ to be (correctly) rejected with probability 0.84. In the design with conditional heteroskedasticity the SNIV confidence set is marginally wider than under homoskedasticity. Each optimization problem is solved more slowly than under the classical design, taking around 36 seconds on average. Nevertheless, the problems are sufficiently tractable so as to allow informative inference on a standard machine.

%In sum, using semidefinite hierarchies allows us to compare the SNIV and AR confidence sets in the computationally challenging setting of $d_X=10$ endogenous regressors, for which a grid search is impossible. For the designs we consider, the SNIV confidence set performs similarly to the AR confidence set in classical settings, but can also be informative in settings with many, weak, invalid and endogenous instruments.

\section{Conclusion}\label{conc}

We use self-normalization of sample moments to conduct robust, computationally tractable inference in linear instrumental variables models. We also show that our computational approach is not unique to self-normalzation, and can be applied to perform fast inversion of other tests. In our view there are two avenues for future work. 

First, though SNIV requires minimal assumptions and has desirable statistical and computational properties, when $d_X$ is large it can be conservative when the object of interest is low dimensional (e.g., a single treatment effect). Though this is not an issue in the leading case in which $d_X$ is small (e.g., one endogenous regressor, possibly after partialing our a small number of exogenous regressors) and $d_Z$ may be large (with possibly weak instruments), future work may seek to adapt our approach to perform robust inference directly on a \textit{sub-vector} of parameters of interest. 

Second, we believe that our computational approach is applicable beyond the instrumental variables context. An obvious setting to which our results may be applied is that of inference in partially identified models, which is often based on solving programming problems such as \eqref{prob}. A simple example in which the optimization problem is semi-algebraic (hence our approach is applicable) is the $2\times2$ entry game considered by \cite{kaido19}.

\bibliographystyle{ier}
\bibliography{refs}

\section*{Appendix}

\noindent{\bf Proof of Proposition \ref{propcover}.} The result follows by applying a union bound to the bounds in \cite{jing03} (Class 1), \cite{bertail08} (Class 2) or \cite{pinelis94} (Class 3), which yield the corresponding values of $r_n$ in the main text. For Class 1, the coverage is asymptotic because $C_1\mu_{2+\delta}\left(1+\sqrt{n}r_n\right)^{2+\delta}n^{-\delta/2}\to 0$ where $C_1$ is an unknown universal constant.  For Class 2, the results of \cite{bertail08} yield the bound $\sqrt{2/(n/\log(d_Z(2e+1)/\alpha)-\mu_4)}$ and we use  $n-\mu_4\log({d_Z}(2e+1)/\alpha)\geq n/2$ to obtain $r_n$ which does not depend on the unknown $\mu_4$.  \hfill$\square$\\

\noindent{\bf Proof of Proposition \ref{propcovermboot}.} The result follows from Corollary 2.1 in \cite{chernozhukov13} and the fact that $\E_n[U(\beta)^2]$ is consistent for $\E[U(\beta)^2]$ under the conditions of Class 4.  \hfill$\square$\\

\noindent{\bf Proof of Proposition \ref{propsnivg}.} Under Classes 1-3, the first inequality in the definition of the SNIV confidence set can be rewritten as
\begin{align}
\E_n[(Z_lU(\beta)-\theta_l)^2]^{-1/2}|\E_n[Z_lU(\beta)]-\theta_l|\leq r_n\quad \forall l\in[d_Z].
\end{align}
Squaring both sides and rearranging yields the equivalent degree 2 polynomial inequalities
\begin{align}
r_n^2\E_n[(Z_lU(\beta)-\theta_l)^2]-(\E_n[Z_lU(\beta)]-\theta_l)^2\geq0\quad \forall l\in[d_Z].
\end{align}
Under Class 4 (which is not applicable with potentially endogenous instruments), we obtain instead the degree 2 polynomial inequalities
\begin{align}
 \widehat{r}^2\E_n[Z_l^2]\E_n[U(\beta)^2]-\E_n[Z_lU(\beta)]^2\geq 0\quad \forall l\in[d_Z].
\end{align}
Without loss of generality, suppose that we order the indices of the regressors such that $S_Q=[d_Q]$. The second inequality in the definition of the SNIV confidence set is equivalently expressed using the polynomial (in)equalities
\begin{align}
\exists \zeta\in[0,1]^{d_Q}:\quad 	&\zeta_k^a(1-\zeta_k)^b=0&\forall k\in S_Q,(a,b)\in\mathbb{N}^2,\nonumber \\
					&(1-\zeta_k)^a\beta_k=0&\forall k\in S_Q,a\in\mathbb{N},\nonumber\\
					&s-\sum_{k\in S_Q}\zeta_k\geq 0,
\end{align}
where, due to the constraint $\zeta_k(1-\zeta_k)=0$, $\zeta$ comprises $d_Q$ indicators for the nonzero entries of $\beta_{S_Q}$ (see \cite{feng13}). The third inequality in the definition of the SNIV confidence set is obtained identically introducing $\eta\in[0,1]^{\widetilde{d}_Q}$. Since equalities can be defined using two inequalities of opposing directions we can stack all of the polynomial inequalities as $\widehat{\mathbf{g}}(\beta,\theta,\gamma)\geq 0$ where $\gamma\triangleq(\zeta^\top,\eta^\top)^\top$. \hfill$\square$\\
\newpage
\FloatBarrier
\marginsize{1cm}{1cm}{0.25cm}{0.25cm}
\pagenumbering{gobble}

\begin{center}
\begin{table}[t!]
\caption{Monte Carlo}\label{tab:sims}
{\footnotesize
\begin{tabular}{lccccccccccc}
\hline\hline
\multicolumn{10}{c}{\textbf{Classical} ($d_Z=10,\pi^*=0.3$)}\\
\hline
&$\beta_1$&$\beta_2$&$\beta_3$&$\beta_4$&$\beta_5$&Cover&Exact&Time (s)&$\theta\neq 0$\\
\cmidrule(lr){2-6} \cmidrule(lr){7-7} \cmidrule(lr){8-8}  \cmidrule(lr){9-9}  \cmidrule(lr){10-10} 
\textbf{AR} &0.361 &   0.361  &  0.360  &  0.361  &  0.360&0.942&1.000&1.08\\
&&   &   &  &  &(0.994)\\

\textbf{SNIV} (Class 1) &0.343 &   0.343 &   0.341 &   0.342  &  0.343&0.944&1.000&1.34\\
&&   &   &  &  &(0.992)\\

\textbf{SNIV} (Class 1, het.) &0.349 &   0.349 &   0.347 &   0.348  &  0.349&0.946&1.000&12.10\\
&&   &   &  &  &(0.996)\\

\textbf{SNIV} (Class 3) &0.419 &   0.420 &   0.418 &   0.419  &  0.419&0.988&1.000&15.57\\
&&   &   &  &  &(1.000)\\

\textbf{AR} (CI) &0.163&    0.163 &   0.163 &   0.163  &  0.163&0.949&1.000&1.07\\
\hline

\multicolumn{10}{c}{\textbf{Many instruments} ($d_Z=1999,\pi^*=0.3$)}\\
\hline

\textbf{AR} &19.373  & 19.427  & 19.374  & 19.391  & 19.399&0.324&0.995&0.73\\
&&   &   &  &  &(1.000)\\

\textbf{SNIV} (Class 1) &0.634  &  0.635 &   0.632  &  0.633  &  0.633  &0.956&0.855&4.17\\
&&   &   &  &  &(1.000)\\

\textbf{SNIV}  (Class 1, het.)&0.608  &  0.610 &   0.607  &  0.608  &  0.606  &0.954&0.874&8.29\\
&&   &   &  &  &(1.000)\\

\textbf{SNIV} (Class 3)&0.734  &  0.736 &   0.733  &  0.733  &  0.733  &0.988&0.692&9.48\\
&&   &   &  &  &(1.000)\\

\textbf{AR} (CI) &19.373  & 19.427  & 19.374 &  19.391  & 19.399&1.000&0.993&0.73\\
\hline
\multicolumn{10}{c}{\textbf{Many instruments} ($d_Z=2100,\pi^*=0.3$)}\\
\hline
\textbf{SNIV} (Class 1)&0.635  &  0.632 &   0.635  &  0.632  &  0.634  &0.954&0.863&4.41\\
&&   &   &  &  &(1.000)\\

\textbf{SNIV} (Class 1, het.)&0.609  &  0.608 &   0.611  &  0.608  &  0.609  &0.952&0.878&8.50\\
&&   &   &  &  &(1.000)\\

\textbf{SNIV} (Class 3)&0.735  &  0.732 &   0.736  &  0.732  &  0.735  &0.988&0.681&9.73\\
&&   &   &  &  &(1.000)\\
\hline
\multicolumn{10}{c}{\textbf{Weak instruments} ($d_Z=10,\pi^*=0.03$)}\\
\hline

\textbf{AR}&6.4959   & 6.314  &  6.208   & 6.253  &  6.189&0.942&0.991&1.22\\
&&   &   &  &  &(1.000)\\

\textbf{SNIV} (Class 1)&12.776&   12.833  & 12.782  & 12.892  & 12.875  &0.944&0.999&1.59\\
&&   &   &  &  &(1.000)\\

\textbf{SNIV} (Class 1, het.)&12.771&   12.823  & 12.775  & 12.886  & 12.869  &0.946&1.000&12.48\\
&&   &   &  &  &(1.000)\\

\textbf{SNIV} (Class 3)&13.819&   13.886  & 13.852  & 13.950  & 13.935  &0.988&1.000&15.95\\
&&   &   &  &  &(1.000)\\

\textbf{AR} (CI) &0.7021   & 0.695   & 0.680   & 0.688  &  0.688&0.963&1.000&1.23\\
\hline
\multicolumn{10}{c}{\textbf{Invalid instruments} ($d_Z=9,\pi^*=0.3$)}\\
\hline
\textbf{SNIV} (Class 1, $s=2$)&0.276  &  0.136 &   0.000 &   0.000   & 0.000 &0.968&0.000&17.09\\
&&   &   &  &  &(0.968)\\

\textbf{SNIV} (Class 1, $s=3$)&1.251&    0.163&    0.096&    0.095&    0.096 &0.968&0.000&16.51\\
&&   &   &  &  &(0.968)\\

\textbf{SNIV} (Class 1, het., $s=2$)&0.280  &  0.220 &   0.000 &   0.000   & 0.000 &0.972&0.000&25.26\\
&&   &   &  &  &(0.972)\\

\textbf{SNIV} (Class 1, het., $s=3$)&1.253  &  0.245 &   0.100 &   0.099   & 0.099 &0.972&0.000&25.05\\
&&   &   &  &  &(0.972)\\

\textbf{SNIV} (Class 3, $s=2$)&0.335  &  0.157 &   0.000 &   0.000   & 0.000 &0.988&0.000&27.21\\
&&   &   &  &  &(0.988)\\

\textbf{SNIV} (Class 3, $s=3$)&1.279  &  0.188 &   0.111 &   0.110   & 0.111 &0.972&0.000&26.64\\
&&   &   &  &  &(0.988)\\
\hline
\multicolumn{10}{c}{\textbf{Endogenous instruments} ($d_Z=12,\pi^*=0.3$)} \\
\hline
\textbf{SNIV} (Class 1, $\widetilde{s}=2$)&0.946   & 0.958 &   1.238   & 1.236  &  1.225  &0.948&0.000&36.01&0.840\\
&&   &   &  &  &(0.976)&&& (0.304)\\

\textbf{SNIV} (Class 1, het., $\widetilde{s}=2$)&0.956   & 0.959 &   1.237   & 1.253  &  1.236  &0.956&0.000&36.42&0.824\\
&&   &   &  &  &(0.992)&&& (0.308)\\

\textbf{SNIV} (Class 3, $\widetilde{s}=2$)&1.206   & 1.211 &   1.543   & 1.553  &  1.537  &0.984&0.000&36.83&0.760\\
&&   &   &  &  &(1.000)&&& (0.296)\\
\hline

\end{tabular}\\
}
\scriptsize{\textbf{Notes:} $d_X=10,n=2000,\beta^*=(1,-1,0,...,0)^\top$. We report the mean width of the confidence region for $\beta_1,...,\beta_5$.  `het'  is the design with conditional heteroskedasticity. `Cover' is for $(\beta^\top,\theta^\top)^\top$ for SNIV and $\beta$ for AR. In parentheses, we include the coverage probability of the outer approximation of the confidence set ($f_\mathbf{u}(\beta)\geq f^*_h(\mathbf{u})$ for all $\mathbf{u}\in\mathcal{U}$). For $\mathcal{U}$ we use a grid of 1600 points over the surface of $\ell_2$ ball of radius 1. `Exact' is the fraction of optimization problems solved exactly ($f^*_h=f^*$). `Time (s)' is the mean time taken to solve an optimization problem in seconds. `$\theta\neq 0$` is the fraction of datasets in which an endogenous instrument was detected. In parentheses, both were detected. All programs use $\overline{h}=2,B=100$. 500 replications.} 
\end{table}
\end{center}

\end{document}